# Oscillating planar Hall response from the surface electrons in bulk crystal Sn doped Bi$_{1.1}$Sb$_{0.9}$Te$_2$S

Bin Wu, Xing-Chen Pan, Wenkai Wu, Fucong Fei, Bo Chen, Qianqian Liu, Haijun Bu, Lu Cao[1*], Fengqi Song*, Baigeng Wang*

*National Laboratory of Solid State Microstructures, Collaborative Innovation Center of Advanced Microstructures, and School of Physics, Nanjing University,*

*Nanjing, 210093, China*

---

[*] The corresponding author. email: caolu@nju.edu.cn, songfengqi@nju.edu.cn, bgwang@nju.edu.cn; Phone: +86-25-83592701

**Abstract**

We report the low-temperature magneto-transport in the bulk-insulating single crystal of topological insulator Sn doped $Bi_{1.1}Sb_{0.9}Te_2S$. The Shubnikov-de Haas oscillations appear with their reciprocal frequency proportional to $\cos\theta$, demonstrating the dominant transport of topological surface states. While the magnetic field is rotating in the sample surface, the planar Hall effect arises with sizeable oscillations following a relation of $\cos\theta\sin\theta$. Its amplitude reaches the maximum at the lowest temperature and drops to nearly zero at the temperature higher than 100 K. All these evidences consolidate such planar Hall oscillations as a new golden criterion on the topological surface transport.

Three-dimensional topological insulators (TIs) have gained much attention these years due to the massless Dirac fermions and the potential application in the spintronics and quantum devices.[1–5] The gapless metallic states, which are protected by the time-reversal symmetry, have been demonstrated on the surface of a series of materials. The transport of the topological surface states (TSS) has been heavily studied with peculiar behaviors in weak anti-localization,[6–8] Aharonov-Bohm interference,[9–11] universal conductance fluctuations,[12–14] Shubnikov-de Haas (SdH) oscillations,[15–17] quantum Hall responses,[18–21] etc., each of which arouses a new subfield pointing to the quantum device applications.

Planar Hall effect (PHE) has been studied in the transport studies of topological materials.[22,23] It was first noticed in the topological semimetals when the community met difficulty in demonstrating the chiral anomaly by longitudinal negative magnetoresistance (MR).[24,25] The measurement of PHE may further support the topological electrons by the unique angle dependence of Hall resistance. PHE is caused by the chiral anomaly in the topological semimetals and has a period of 180 degrees that shows extrema at 45° and 135°.[26–29] Recently, planar hall response was also observed in TI $Bi_{2-x}Sb_xTe_3$ films and proposed to be attributed to the spin–momentum locking feature of the TSS.[30] In addition, the planar hall response is further related to the effect of nontrivial Berry phase and magnetic moments.[31] This might contribute a new golden evidence on the TSS transport.

However, further evidence is still required on the TSS origin of TI's PHE since the thin device may exhibit inter-surface coupling. The complex device fabricating process and the influence of substrate may bring uncertain factors of the transport signals. The planar hall response will be more convincing if it is observed in bulk TI crystals and follows the similar behavior. Recent progress on the single crystal growth shed light on this topic. Many topological materials, with high-quality single crystals, may provide an opportunity to observe the PHE.[32–35] Here we grew the high-quality single crystals of Sn doped $Bi_{1.1}Sb_{0.9}Te_2S$ (Sn-BSTS) with highly insulating bulk state and a prominent Dirac cone in its band structure. We successfully observed the planar Hall effect in such bulk samples.

High-quality $Sn_{0.02}Bi_{1.1}Sb_{0.9}Te_2S$ single crystals were prepared by the melt-grown method.[35,36] The chemical constituents of the Sn-BSTS crystals were confirmed by the energy-dispersive spectroscopy (EDS) and the ratio between Bi, Sb, Te and S was close to 1.1:0.9:2:1. The exact proportion of Sn was 0.03 that was obtained by X-ray fluorescence spectroscopy (XRF). The details for crystal growth, EDS and XRF can be seen in the supplementary materials. Fig. 1(a) shows the EDS mapping of the four elements Bi, Sb, Te and S. It is clearly that the four elements are uniformly distributed. The powder X-ray diffraction pattern of the sample is shown in Fig. 1(b). We refined the experimental data based on a rhombohedral cell. The fitting curve agrees well with the experimental data and the obtained lattice constants, a = 4.21 Å; c = 29.57 Å, is consistent with early experiments.[33,35]

Fig. 1(c) shows the band dispersion of the Sn-BSTS at 10 K characterized by Angle resolved photoemission spectroscopy (ARPES). The Dirac cone with the linear dispersion is obviously seen in the center of the Brillouin zone. Clear surface state with Dirac cone dispersions can be identified and the Dirac point is about 126 meV below the Fermi surface, indicating that the Fermi surface of Sn-BSTS is dominated by the TSS. The temperature-dependent electrical resistivity plot is shown in Fig. 1(d). The resistivity increases as the temperature goes down when over 100 K, which shows an insulator behavior known as the contribution of the bulk state. When below 100 K, the resistivity begins to diminish with the decreasing temperature, indicating a dominated transport by the metallic surface states.[37,38] Large bulk resistivity reaches 14 Ωcm at 2 K.

For further investigating the nature of the topological surface states in Sn-BSTS, we measure the magnetoresistance of our samples at low temperature. As shown in Fig. 2(a), the resistance $R_{xx}$ exhibits obvious Shubnikov-de Hass (SdH) oscillations above 5 T. Fig. 2(b) is the oscillations component of $\Delta G_{xx}$ versus $1/B$ under different temperatures obtained by background subtracting. We can extract an oscillating mode with the frequency of 41.15 T via performing the fast Fourier transformation (FFT) operation. Using the Onsager relation $F = (\hbar/2\pi e)A_F$, the Fermi wave vector of $k_F$ = 0.0354 Å$^{-1}$ and the corresponding surface carrier density of $n_s$ = 9.95 × 10$^{11}$cm$^{-2}$ can be extracted. It is well known in topological insulators that the Berry phase can be affirmed by the intercept of Landau-level fan diagram.[17,39] We plotted the Landau-level fan diagram in Fig. 2(c). The minima of $\Delta G_{xx}$ are treated as the integer indices $n$ and the maxima are plotted as the half integer indices $n + \frac{1}{2}$. The intercept value of the fitting plot is 0.68, which is close to 0.5, indicating the nontrivial Berry phase. The topological surface states contribution of the SdH oscillation thus can be confirmed.

We can obtain other transport parameters by fitting the SdH oscillations to the Lifshitz-Kosevich (LK) formula.[15,17,40] All parameters extracted from the SdH oscillations such as Fermi velocity $v_F$, relaxation time $\tau$ and mean-free path $l_s$ are list in TABLE I. The surface sheet conductance can be calculated by the formula $G_s = (e^2/h)k_F l_s = 2.6 \times 10^{-4} \Omega^{-1}$.[40] The ratio of the surface conductance accounts for 15% of the total conductance at 2 K. This implies that the surface conductance of our Sn-BSTS bulk crystal has the same order of magnitude as the total conductance. Fig. 2(d) shows the normal out-of-plane angle-dependent resistance measurement, measured as rotating the magnetic field in the plane formed by the current and the c axis and when $\theta = 0$, the applied field is perpendicular to the sample surface. The SdH oscillations are clear from 0 to 50 degrees but their amplitudes gradually decrease as the angle increases. We extracted the oscillation frequency through the FFT operation. As shown in the inset of Fig. 2(d), the reciprocal of the oscillation frequency is proportional to $\cos\theta$, demonstrating a strict two dimensional (2D) behavior of the oscillations contributed by the TSS.[15,17] The significant transport contribution from TSS provides an opportunity to investigate the planar Hall effect in the bulk topological insulator Sn-BSTS.

In the next step, we change the field direction and measure the magneto-transport with an in-plane magnetic field rotating parallel to the sample surface as seen by the inset sketch in Fig. 3(a). $\theta$ is the angle between the magnetic field and the direction of longitudinal current. Fig. 3(a), (b) display the longitudinal magneto-resistance and the planar hall resistance measured at 2 K in the magnetic field of 10 T, respectively. The anisotropic longitudinal magneto-resistance (AMR) with the rotating of in-plane magnetic field presents the $-\cos^2\theta$ angle dependence with a period of 180 degrees. However, it is hard to distinguish whether this kind of signal is truly comes from the rotating in-plane field or just the traditional longitudinal magneto-resistance caused by the out-of-plane magnetic field component as the slightly misalignment of the sample is unavoidable. [30]

On the contrary, the planar hall resistance in Fig. 3(b) can be identified as the intrinsic signal comes from the in-plane magnetic field. Though the raw data of PHE (Fig. S2(a) in the supplementary materials) are also out of shape caused by the misalignment, the intrinsic signals can be extracted by specific data processing. Firstly, the ordinary hall resistance induced by the out-of-plane field component can be excluded by summing the raw data of positive and negative magnetic field because it is antisymmetric to the magnetic field.[27,30] Secondly, we deal with data by the formula: $R_{yx} = (R_{yx}(\theta) - R_{yx}(\pi - \theta))/2$ to exclude the effect of longitudinal resistance signals. Because the AMR and ordinary magnetoresistance follow $\cos^2\theta$ angle dependence and are symmetrical with $\theta = 180°$ while planar hall resistance follows $\cos\theta\sin\theta$ angle dependence and is anti-symmetric with $\theta = 180°$. As shown in Fig. 3(b), the extrema peak of the intrinsic planar hall resistance is at $45°$ and $135°$ and the period of 180 degrees is clearly displayed. The red line in Fig. 3(b) is the fit of formula $R_{yx} = \Delta R \cos\theta\sin\theta$. Though it slightly deviated from the experimental data because of the amplitude modulation caused by SdH oscillation, the period of 180 degrees and the symmetrical property of the intrinsic planar Hall effect can be solidly confirmed. [41]

We further investigate the magnetic field and temperature dependence of PHE. Fig. 4(a) is the PHE versus angle $\theta$ at different magnetic fields. It is clear that the amplitude increases linearly with increasing magnetic field. We plotted the amplitude of PHE with magnetic field in Fig. 4(b) and the red line was the linear fit. The fit is slightly offset from the coordinate origin at low field. Fig. 4(c) shows the PHE with a magnetic field of 10T from 2 K to 150 K and Fig. 4(d) displays the corresponding amplitude at different temperature. The amplitude decreases from 2 K to 30 K and turns to increase from 30 K to 60 K. The amplitude begins to decrease again when the temperature goes above 60 K. Obviously, we noted that the amplitude at 150 K was dramatically decreasing, which reduced from 5.7 Ω at 100 K to 0.4 Ω at 150 K, a decrease of about 14 times. Meanwhile, the decrease of ordinary magnetoresistance and hall resistance when the magnetic field was perpendicular to the sample surface reduced only by half in the same temperature range. The origin of PHE is interpreted as a breaking of time-reversal symmetry, which causes the difference of backscattering forbiddance induced by the in-plane magnetic field in Dirac fermions of various spins.[30] According to the temperature-dependent resistivity of our sample (Fig. 1(d)), there is an inflection point at 100 K in the curve. This means the bulk state dominates the conductance above 100 K and the metallic TSS show up below 100 K. The surface states transport is seriously suppressed over 100 K. Thus, the drop at 150K in PHE amplitude can be explained as the suppression of topological surface states. Moreover, according to the raw data of 2 K and 150 K (Fig. S2 in the supplementary materials), the planar resistance holds a period of 180 degrees at 2 K while the period changes to 360 degrees at 150 K, indicating the suppress of PHE signals at higher temperature and the traditional out-of-plane Hall signals dominates at 150 K. These evidences demonstrate that PHE is the result of topological surface states.

In summary, we successfully measured the planar Hall effect in bulk single crystals of the topological insulator Sn-BSTS. It presented the angle dependence of $\cos\theta\sin\theta$ and had a nearly linear amplitude increase with the increase of magnetic field. Through the analysis of the temperature dependence of the PHE, we observed an obvious decrease of its amplitude at 150 K that confirmed the PHE is caused by the topological surface states. PHE will provide a new gold criterion to study the topological surface states and have promising applications in quantum devices.

See supplementary materials for details on the crystal growth, EDS spectrum, XRF results and raw data of the planar hall resistance.

**Acknowledgements**

We gratefully acknowledge the financial support of the National Key R&D Program of China (2017YFA0303203), National Key Projects for Basic Research of China (2013CB922103), the National Natural Science Foundation of China (U1732273, U1732159, 91421109, 91622115, 11522432), the Fundamental Research Funds for the Central Universities, and the opening Project of Wuhan National High Magnetic Field Center. The technical support from the Hefei National Synchrotron Radiation Laboratory is also acknowledged.

(2012).

[41]S. Liang, J. Lin, S. Kushwaha, R.J. Cava, and N.P. Ong, preprint arXiv:1802.01544 (2018)

**Figure Captions**

**FIG. 1. The characterization of the Sn-BSTS single crystals**. (a) The EDS mapping of Bi, Sb, Te and S, respectively. (b) The powder XRD pattern of crushed single crystals. The red curve shows the raw data and the black line is the fitting curve. The blue curve is the difference between raw data and the fitting curve and green rods mark the positions of the Bragg peaks. (c) The electronic structure measured by the ARPES. (d) The temperature dependence of electrical resistivity plot.

**FIG. 2. The SdH oscillations of the Sn-BSTS crystal.** (a) The magnetoresistance at different temperatures when the applied field is perpendicular to the sample surface. (b) The extracted SdH oscillations of conductance versus $1/B$. (c) The Landau-level fan diagram. Square dots and round dots represent minima and maxima of the SdH oscillation, respectively. (d) The SdH oscillations measured at different angles with a step of 10 degrees. The applied magnetic field is rotating within the plane of the current and c axis. $\theta$ is the angle between the field and c axis. The inset shows the angle dependence of the SdH oscillations frequency.

**FIG. 3. The anisotropic magnetoresistance and planar Hall resistance.** (a) The longitudinal anisotropic magnetoresistance at 2 K, 10 T. The inset shows the sketch for the measurements of PHE. The magnetic field is rotating in the sample surface and $\theta$ is the angle between the field and the current. (b) The PHE observed at 2 K, 10 T. The red solid

line is the best fit to the formula.

**FIG. 4. The evolution and analysis of the PHE.** (a) The PHE versus angle $\theta$ at different magnetic field. (b) The magnetic-field dependence of the amplitudes of PHE and the linear fitting. (c) The PHE versus angle $\theta$ under different temperatures. (d) The temperature dependence of the amplitudes of PHE. The inset is the sketch of the difference of backscattering forbiddance in Dirac fermions. Red arrows mean the spins of Dirac fermions and gray arrow is the magnetic field.

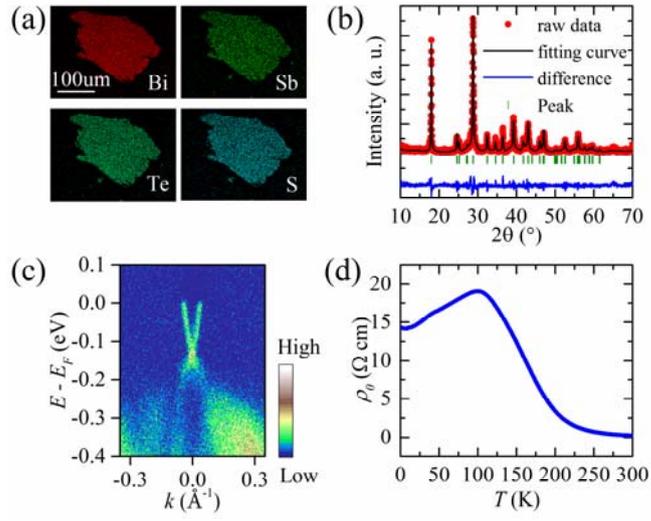

**FIG. 1. The characterization of the Sn-BSTS single crystals.**

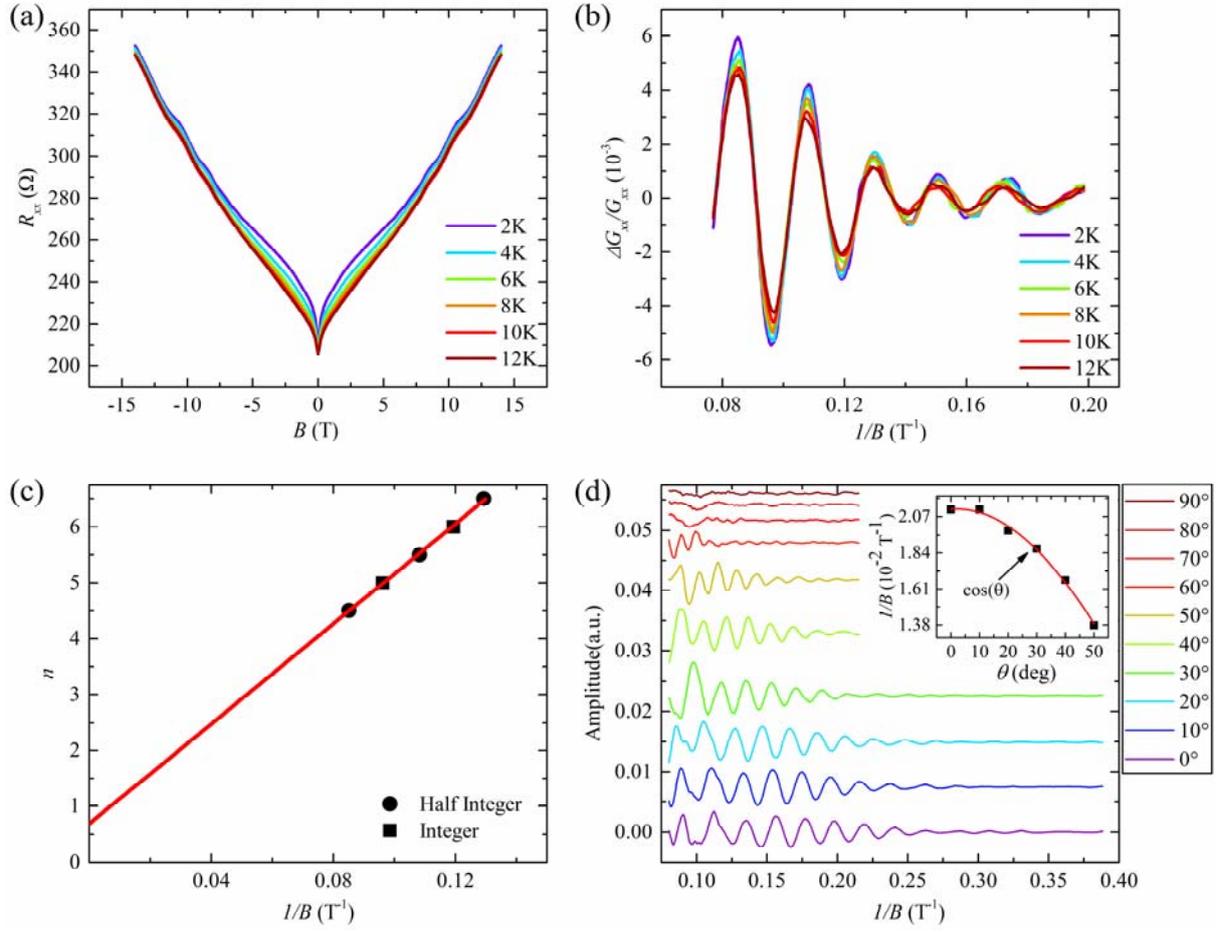

**FIG. 2. The SdH oscillations of the Sn-BSTS crystal.**

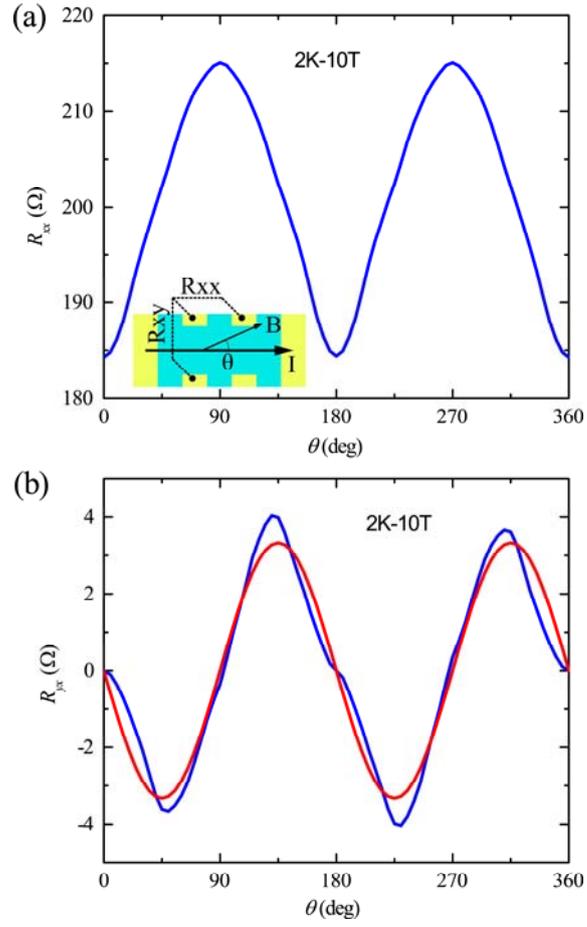

**FIG. 3. The anisotropic magnetoresistance and planar Hall resistance.**

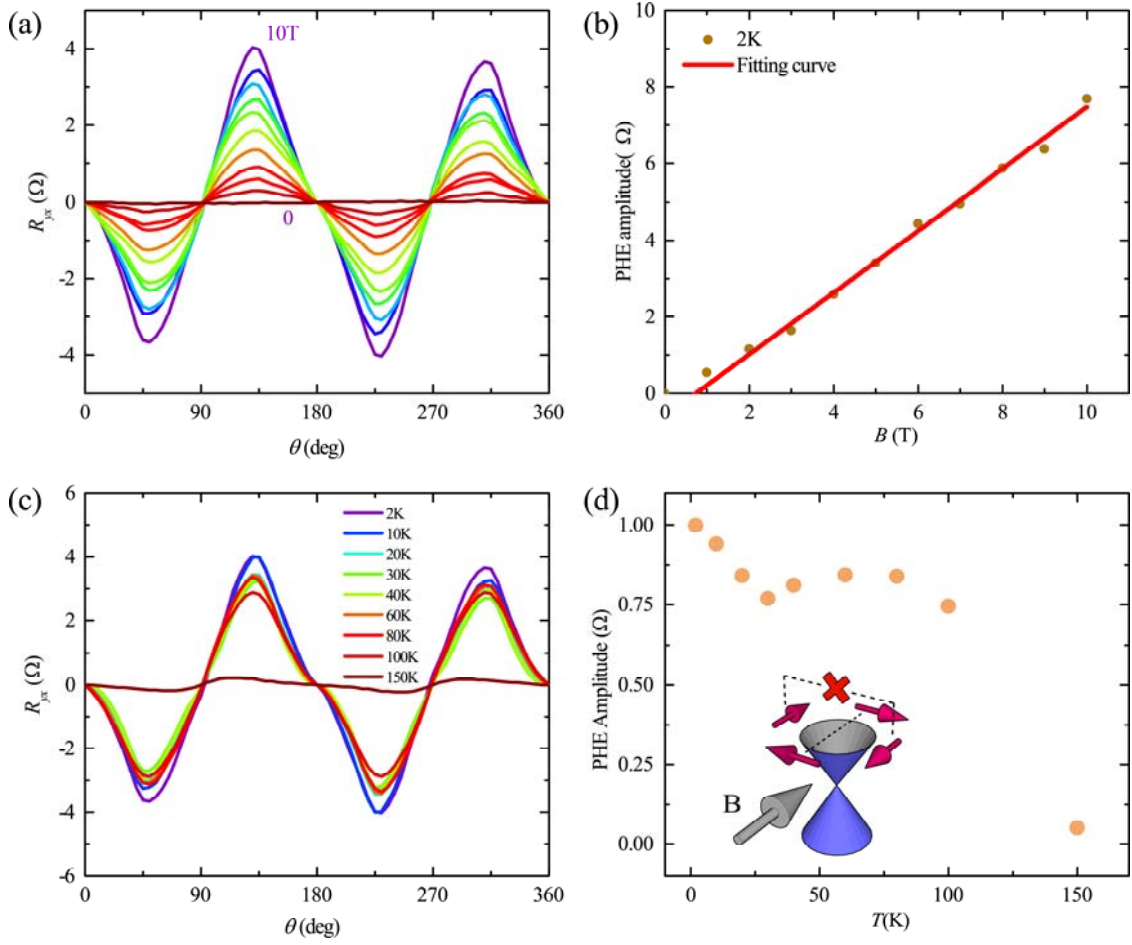

**FIG. 4. The evolution and analysis of the PHE.**

TABLE I. transport parameters extracted from the SdH oscillations.

| $F$ (T) | $k_F$ (Å$^{-1}$) | $n_s$ (cm$^{-2}$) | m* ($m_e$) | $v_F$ (10$^6$m/s) | $T_D$ (K) | $\tau$ (10$^{-14}$s) | $l_s$ (nm) | $\mu$ (cm$^2$V$^{-1}$s$^{-1}$) |
|---|---|---|---|---|---|---|---|---|
| 41.15 | 0.0354 | 9.95×10$^{11}$ | 0.074 | 0.553 | 35.4 | 3.43 | 19 | 820 |